\documentstyle[12pt]{article}
\def\a{\alpha}
\def\b{\beta}

\def\d{\delta}
\def\e{\epsilon}

\def\g{\gamma}

\def\k{\kappa}
\def\l{\lambda}

\def\s{\sigma}

\def\G{\Gamma}

\def\O{\Omega}

\def\S{\Sigma}

\def\inbar{\vrule height1.5ex width.4pt depth0pt}
\def\rlx{\relax\leavevmode}
\def\I{\leavevmode\hbox{\small1\kern-3.8pt\normalsize1}}
\def\openone{\leavevmode\hbox{\small1\kern-3.3pt\normalsize1}}
\def\Ione{\rlx{\rm 1\kern-2.7pt l}}
             \font\cmss=cmss10
             \font\cmsss=cmss10 at 7pt
\def\ZZ{\rlx\leavevmode
             \ifmmode\mathchoice
                    {\hbox{\cmss Z\kern-.4em Z}}
                    {\hbox{\cmss Z\kern-.4em Z}}
                    {\lower.9pt\hbox{\cmsss Z\kern-.36em Z}}
                    {\lower1.2pt\hbox{\cmsss Z\kern-.36em Z}}
             \else{\cmss Z\kern-.4em Z}\fi}
\def\Ik{\rlx{\rm I\kern-.18em k}}  
\def\IC{\rlx\leavevmode
             \ifmmode\mathchoice
                    {\hbox{\kern.33em\inbar\kern-.3em{\rm C}}}
                    {\hbox{\kern.33em\inbar\kern-.3em{\rm C}}}
                    {\hbox{\kern.28em\sinbar\kern-.25em{\rm C}}}
                    {\hbox{\kern.25em\ssinbar\kern-.22em{\rm C}}}
             \else{\hbox{\kern.3em\inbar\kern-.3em{\rm C}}}\fi}
\def\IP{\rlx{\rm I\kern-.18em P}}
\def\IR{\rlx{\rm I\kern-.18em R}}
\def\IN{\rlx{\rm I\kern-.20em N}}

\newcommand{\ol}\overline
\newcommand{\ti}\tilde
\newcommand{\wt}\widetilde
\newcommand{\wh}\widehat
\newcommand{\bv}\breve
\newcommand{\dg}\dagger

\newcommand{\be}{\begin{equation}}
\newcommand{\ee}{\end{equation}}
\newcommand{\bl}{\begin{eqnarray}&}
\newcommand{\el}{&\end{eqnarray}}
\newcommand{\bq}{\begin{eqnarray}}
\newcommand{\eq}{\end{eqnarray}}

\renewcommand{\thefootnote}{\fnsymbol{footnote}}
\newpage
\begin{document}
\begin{center}
{\LARGE{\bf Matrix-Spacetimes and a $2D$ Lorentz-}}\\
{\quad}\\
{\LARGE{\bf Covariant Calculus in Any Even Dimension.}}\\
{\quad}\\
{\large L.P. Colatto$^{(a\star)}$, M.A. De Andrade$^{(b\star)}$ 
and F. Toppan$^{(c\star)}$}
~\\
\quad \\
{\em $\quad^{(a)}$ (UERJ, Instituto de F\'\i sica DFT, 
Rua S\~ao Francisco Xavier, 524, Rio de Janeiro (RJ),  Brasil)}\quad\\
{\em $\quad^{(b)}$ (UCP, F\'\i sica Te\'orica, Rua Bar\~{a}o do
Amazonas, 124, cep 25685-070, Petr\'opolis (RJ), Brasil)}\quad\\
{\em $\quad^{(c)}$ (UFES, CCE Departamento de F\'{\i}sica,
Goiabeiras cep 99060-900, Vit\'{o}ria (ES), Brasil)} 
\quad \\
{\em $\quad^{(\star)}$ (CBPF, DCP Rua Dr. Xavier Sigaud 150, cep 
22290-180 Rio de Janeiro
(RJ), Brasil)} 
\end{center}
{\quad}\\
{\quad}\\
\centerline{\large {\bf Abstract}}

\quad \par
A manifestly Lorentz-covariant calculus based on two matrix-coordinates
and their associated derivatives is introduced. It allows formulating 
relativistic field theories in any even-dimensional spacetime.
The construction extends a single-coordinate matrix formalism based
on coupling spacetime coordinates with the corresponding $\G$-matrices.\par 
A $2D$ matrix-calculus can be introduced for each one of the structures, 
adjoint, complex and transposed acting on $\G$-matrices.
The adjoint structure works for spacetimes with $(n,n)$ signature only. 
The complex structure requires an even number of timelike directions. 
The transposed structure is always defined. A further structure 
which can be referred as ``spacetime-splitting" is based on a fractal 
property of the $\G$-matrices. It is present in spacetimes with dimension 
$D=4n+2$.\par
The conformal invariance in the matrix-approach is analyzed. 
A complex conjugation is present for the complex structure, therefore 
in euclidean spaces, or spacetimes with $(2,2)$, $(2,4)$ signature and so 
on.\par  
As a byproduct it is here introduced an index which labels the classes of 
inequivalent $\G$-structures under conjugation performed by real and 
orthogonal matrices. At least two timelike directions are necessary to get 
more than one classes of equivalence. Furthermore an algorithm is presented 
for iteratively computing $D$-dimensional $\G$ matrices from the $p$ and 
$q$-dimensional ones where $D=p+q+2$.\par
Possible applications of the $2D$ matrix calculus concern the investigation 
of higher-dimensional field theories with techniques borrowed from 
$2D$-physics. 
\\
{\quad}\\
\noindent

\\
{\quad}\\
\rightline{CBPF-NF-063/98\quad\quad\quad\quad\quad\quad}
{\em E-Mails:}\\ 
{\em colatto@cat.cbpf.br}\\
{\em marco@cat.cbpf.br}\\
{\em toppan@cat.cbpf.br}
\newpage
\pagestyle{plain}
\renewcommand{\thefootnote}{\arabic{footnote}}
\setcounter{footnote}{0}

\noindent{\section{Introduction.}}

\par
In the last few years physicists started getting accustomed 
with the likely possibility that the ultimate theory would be
non-commutative. Many investigations on the role of non-commutative 
geometry took place \cite{noncom}. 
In a somewhat different context, 
attempts to penetrate the mysteries of M-theory have been made 
invoking the so-called M(atrix) theory \cite{mtheory} 
and matrix 
string theory
\cite{mstring}. 
The latter in particular is a non-perturbative 
formulation
which allows a non-trivial dynamics for strings by assuming the
target-space coordinates being of matrix type.\par
The above-mentioned approaches are not immediately related with the 
topics discussed in the present paper but they constitute their natural
premises and background. Moreover the results here discussed point to 
further investigations in that direction.\par
Our present work deals with the issue of finding a 
manifestly Lorentz-covariant description of relativistic 
field theories in any even spacetime dimension in terms of a formalism 
which involves matrix-type coordinates. Let us postpone for a 
while answering the question why should we bother about such a 
formulation and let us first discuss the main ideas involved.\par
It is somewhat a trivial remark, found in standard 
textbooks \cite{barut},
that the Lorentz-group can be recovered and interpreted in terms of
matrix-type coordinates. On the other hand it is clear,
following the original ideas of Dirac, i.e. expressing the 
d'Alembertian
$\Box$ box operator through its squared root $\slash{\! \! \! \partial}$, 
that to such a derivative can be associated, as in  any case involving 
derivatives, a
space expressed through a coordinate which is now 
matrix-valued. A Lorentz-covariant calculus, endorsed with 
matrix-type
integrals is immediately at disposal. The above considerations are
perhaps not very deep. In any case they did not find applications 
especially 
because they lead to feasible descriptions, but nothing is gained
and much is lost with respect to the standard case. 
The reason is clear, the lack of extra-structures. For instance, 
if we work in
a single coordinate matrix-type formalism, then we have no room left
to introduce in our theories antisymmetric tensors like 
curvatures $F_{\mu \nu}$ which
require antisymmetry properties among indices and therefore at 
least two coordinates. The restriction is so strong that we are 
not even
allowed to formulate QED or Yang-Mills theories. Therefore, if we wish 
to play the
 ``matrix game" in a purposeful way
we need at least two coordinates.\par
In reality ``two" is quite sufficient for our scopes. More than that,
it is precisely what we need. Indeed two dimensions are just 
enough to formulate all kind of theories we could be 
possibly interested in. Besides, an impressive list of methods and 
techniques have been elaborated to deal
precisely with field theories in $2D$. Let us just mention one 
issue for all, integrability. Integrable field theories 
are well understood in $2D$ \cite{lax} due to the possibility of 
representing 
equations of motion as zero-curvature equations in the form 
$[\partial_Z - L_Z, \partial_{\overline Z} -L_{\overline Z}] =0$ 
where $L_Z$, $L_{\overline Z}$ are Lax pairs. In
higher dimensions analyzing integrability is much more
problematic \cite{integrability}.
We have reasons to believe that our 
approach could 
shed light on this subject. 
Indeed the point of view we are advocating here is that we can, 
formally, deal even-dimensional
spacetimes as a matrix-valued $2D$ space. With a pictorial image, 
we can say that we boost
dimensions to the Flatland. \par
It is clear that 
non-commutative features are present with 
respect to theories formulated on the plane. These extra structures
however, far from being undesired, are welcome and natural.
They are the expected price we must pay for living in a 
higher-dimensional world.   \par
Even if as a consequence we are not automatically guaranteed 
that working methods in
the standard $2D$-plane continue to work in high-$D$, 
nevertheless our approach helps 
attacking problems with techniques which, so to speak, 
are ``driven by the $2D$ formalism 
itself".  
As an example and with respect to the above mentioned integrability 
issue,
this would imply investigating the 
matrix-analogs of the ordinary $2D$ Lax pairs. 
A (partial) list of other topics and areas which could benefit 
from this approach will be discussed in the
conclusions.\par 
The key ingredient we demand is acting with even-dimensional 
Poincar\'{e}
generators on two matrix-valued coordinates $Z, {\overline Z}$ which, 
in order to leave the
construction as simple as possible, we require being mutually 
commuting 
\begin{eqnarray}
\relax {\em i)}&& [Z, {\overline Z}] =0
\label{comm}
\end{eqnarray}
The differential calculus involves two derivatives $\partial_Z$,
$\partial_{\overline Z}$ which 
should satisfy a factorization (Lorentz-covariant) property as
follows
\begin{eqnarray}
\relax {\em ii)} && \partial_Z\partial_{\overline Z} \propto \Box \cdot\I
\end{eqnarray}
This property can also be rephrased in more geometrical terms by
requiring the (pseudo)-euclidean quadratic form 
$ds^2=dx_\mu dx_\nu\eta^{\mu\nu}$ being expressed through 
$d{\overline Z}\cdot dZ = ds^2\cdot \I$, where $dZ$, $d{\overline Z}$
are matrix-valued differentials. The commutativity of $Z, {\overline Z}$
implies the commutativity of the derivatives, therefore 
$\partial_Z\partial_{\overline Z}= \partial_{\overline Z}\partial_Z$.\par
The disentangling of $Z,{\overline Z}$ further requires that
\begin{eqnarray}
{\em iii \quad a)} 
&& \partial_Z {\overline Z} = \partial_{\overline Z} Z =0
\end{eqnarray}
while the normalization condition 
\begin{eqnarray}
{\em iii \quad b)} 
&&
\partial_Z Z =\partial_{\overline Z}{\overline Z} = \I
\end{eqnarray}
can be imposed. Please notice that in the above formulas the action 
of derivatives is a left action (not a free one).
\par
The three listed properties are non-trivial ones. In 
order to make them work two different schemes can be adopted. 
The first one is based on non-trivial identities satisfied by the
Clifford 
$\Gamma$-matrices and involving vector-indices contractions
(from time to time we refer to such identities as ``vector-traces", 
being understood they are not the standard traces
taken w.r.t the spinorial indices). 
The second one uses a fractal property of the same 
$\Gamma$-matrices, i.e. an
algorithm which allows computing higher-dimensional $\Gamma$-matrices from
lower-dimensional ones. As we discuss later in the text, this 
fractal property encodes the information that the Lorentz-algebra has 
the structure of a homogeneous space.\par
The ``vector trace"-case should be analyzed for each one of the three 
structures, adjoint, transposed
or complex, which act on $\Gamma$ matrices. While the transposed 
structure allows to satisfy the three
properties above for any even 
space-time, 
the adjoint action restricts the $D= 2n$ spacetime to have $(n,n)$ 
signature, 
and the complex structure restricts the signature to have 
an even number of time-coordinates. In the case of the complex 
structure $Z , {\overline Z}$ 
are mutually complex conjugated (${\overline Z} = Z^{*}$), 
while no conjugation is 
present in all the remaining cases. \par
The second scheme, which for reasons that will become clear later
will be referred as the ``splitting case", works only when the
dimensionality of the spacetime is restricted to the values
$D= 4n+2$ ($n$ is a non-negative integer). In this case the signature
is arbitrary.\par 
Contrary to the standard calculus, the matrix-calculus here discussed 
naturally encodes the mentioned trace or fractal  properties. \par
It is worth mentioning that when the formulas here reported 
(for the whole set of constructions
mentioned above) are specialized to the $D=2$ case, we trivially recover the
ordinary $2D$ formalism in either the euclidean or Minkowski spacetime.\par 
As we discuss at length in the text our calculus is manifestly 
Poincar\'e covariant and a $2D$ matrix-integration can be easily
constructed. We explicitly apply it to bosonic theories and
QED fields to show how to recover the results of the 
standard formulation.\par
It is worth mentioning that as a byproduct of the matrix-construction 
here discussed some other results are found. In particular, motivated
by finding the consistency conditions under which the complex
structure gives rise to a  $2D$-matrix coordinate
calculus, we are able to introduce an index
which labels the classes of inequivalent $\G$-structures under conjugation
realized by matrices both real and orthogonal. This index is shown to
classify the Wick rotations mapping the euclidean $D$-dimensional space to
a spacetime with $(k,D-k)$ signature.\par 
The algorithm mentioned before is here furnished. It is a realization of  
$D$-dimensional $\G$-matrices in terms of $p$ and $q$-dimensional 
ones, where $D,p,q$ are even integer numbers satisfying the relation 
$D=p+q+2$.\par  
The scheme of the paper is the following.\par 
In section $2$ we
introduce and discuss at first the covariant calculus for a single matrix 
coordinate. In section $3$ the conformal invariance is analyzed in the light
of the matrix-approach. In section $4$ the basic properties 
concerning $\G$ matrices, as well as the conventions used, are reported.
The algorithm expressing higher-dimensional $\G$-matrices from 
the lower-dimensional ones is presented in section $5$. 
Section $6$ is devoted to discuss the consistency conditions for a 
$2D$ matrix-calculus in
the ``vector-trace" approach. It is shown that the vanishing of
$\G^\mu {\G_\mu}^\dagger$, $\G^\mu {\G_\mu}^*$ or $\G^\mu {\G_\mu}^T$
is required. In section $7$ the complete solution is furnished. The already
mentioned restrictions to the $2D$ matrix-calculus with the adjoint or
complex structure arise as a consequence. In section 8 the index
discussed before is introduced and computed. It is shown how to 
relate it to Wick rotations from euclidean spaces to pseudoeuclidean 
spacetimes. 
The $2D$ matrix formalism is revisited and compact formulas are given in
section $9$. In section $10$ a relativistic separation of the 
matrix-variables is explained. In section $11$ the formula realizing  
higher-dimensional $\G$-matrices from lower-dimensional ones is used
to present a different (inequivalent) way of introducing the $2D$ 
matrix-coordinate calculus. It applies for $p=q$, that is when the
spacetime is $D=4n+2$-dimensional. Section $12$ is devoted to explain how 
to apply the matrix-calculus to forms. In the conclusions we make some
comments about the $2D$ matrix calculus and discuss its
possible applications.

\vspace{0.2cm}
\noindent{\section{Matrix coordinates.}}

Originally the $\slash{\! \! \! \partial}$ derivative was 
introduced by Dirac to be 
applied on spinors in order to define the dynamics of spinorial 
fields. However, as mentioned in the introduction, 
$\slash{\! \! \! \partial}$
admits another interpretation. Indeed it can be regarded as 
acting on a matrix-valued coordinate space. It
turns out that e.g. bosonic fields can be described within
a Lorentz-covariant framework in such a manner.\par 
Since the idea of
using matrix coordinates is at the very core of our further 
developments
let us introduce and discuss in some detail 
the theory of a single-matrix coordinate at first.\par
We consider the following matrix-valued objects:\par
${\em i)}$ the matrix coordinate $Z= x_{\mu}\Gamma^{\mu}$\par
${\em ii)}$ the matrix derivative \footnote{ the ${\textstyle{1\over D}}$ 
normalization is introduced for convenience in order to
normalize $\partial_Z\cdot Z = \I$.} 
$\partial_Z= {\textstyle{1\over D}}\partial_{\mu}\Gamma^{\mu}$\par
${\em iii)}$ the matrix differential $dZ = dx_{\mu}\Gamma^{\mu}$.\par
The above objects are all $\Gamma$-valued, where the $\Gamma^{\mu}$ 
denote any set of $D$-dimensional 
$\Gamma$-matrices (the signature of the space-time does not play
any role for the moment and can be left arbitrary).\par
Matrix-valued functions are $\Gamma$-valued functions ($\Phi$) of the 
single $Z$ matrix-variable (i.e. $\Phi\equiv \Phi (Z)$). 
Both Lorentz and Poincar\'e invariances are 
automatically encoded in the above formalism. Indeed not only
$\partial_Z^2 = {\textstyle {1\over D^2}} \Box \cdot {\I}$, 
but also the quadratic form
$d Z^2$ satisfies 
\begin{eqnarray}
d Z^2 &=& d s^2 \cdot \I
\label{quadraticform}
\end{eqnarray} 
(here, as in the introduction, $d s^2 = dx_{\mu}dx_{\nu}\eta^{\mu\nu}$). 
\par
It turns out that linear transformations
which include the Poincar\'e group as a subgroup leave invariant
this quadratic form. 
Indeed the differential $d= dx_{\mu}\partial^{\mu}$ can be reexpressed in
matrix-coordinate form as
\begin{eqnarray}
d\cdot \I &=& {{D\over 2}} ({dZ \cdot {\partial\over\partial Z}}
+{\partial\over\partial Z}\cdot dZ)
\end{eqnarray}
so that for a generic $f(Z)$ function of $Z$ we can write 
$df = {\textstyle{D\over 2}} 
(dZ\cdot{\textstyle{ \partial \over\partial Z}}
+{\textstyle{\partial \over\partial Z}}\cdot dZ)\cdot f$. 
Notice that when $f$ is the
identity ($f(Z)\equiv Z$) we recover, as it should be, 
the above definition for $dZ$. We wish to point out that,
since we are dealing with matrix-valued objects, some care
has to be taken when performing computations with respect
to the ordinary case. Non-commutative issues imply for instance
that $dZ\cdot Z \neq Z\cdot dZ$. \par
If we specialize the $f$-transformation to be given by
\begin{eqnarray}
Z' &=& f (Z)= S\cdot Z \cdot S^{-1} + K
\end{eqnarray}
where $S$ is an element of the $D$-dimensional Lorentz group 
(i.e. $S\Gamma^{\mu}S^{-1} = {\Lambda^{\mu}}_{\nu} \Gamma^{\nu}$) 
and $K$ is a constant matrix
which for what we need is sufficent to take of the form
$K = k_{\mu}\cdot \Gamma^{\mu}$, we therefore obtain
$dZ'= S\cdot dZ\cdot S^{-1}$ which further implies 
${dZ'}^2 = {dZ}^2$ since the latter is proportional to the 
identity.\par
The calculus can be further enlarged to accomodate a formal
definition of a matrix-valued volume 
integration form and a matrix-valued delta-function.
They both coincide with the standard manifestly relativistic covariant 
definitions. To express them in matrix form is
sufficent to recall the definition of $\Gamma^{D+1}$, the 
$D$-dimensional analog of $\gamma^5$, as the Lorentz-invariant 
product
of the $D$-dimensional $\Gamma^{\mu}$
\bq
\Gamma^{D+1} &=& \e \G^0\cdot\G^1\cdot ...\cdot \G^{D-1}
\eq
with $\e =(-1)^{{\textstyle{(s-t)\over 4}}}$. Here $t$ denotes the
number of timelike coordinates with $+$ signature and $s= D-t$ the
number of spacelike coordinates with $-$ signature.
Therefore we can write
\bq
dV &=& dx_0\cdot ... \cdot dx_{D-1}\cdot \I = \e d\G(0)\cdot ...d\G(D-1)
\cdot \G^{D+1}
\label{gammafivegen}
\nonumber
\eq
(here $d\G (i) = dx_i \G^i$) and
\bq
\delta (Z,W) &=& \delta(x_0-y_0) \cdot...\cdot
\d (x_{D-1} - y_{D-1}) \cdot \I = \e \d_{\G}(0)\cdot ...\cdot \d_{\G}(D-1)
\nonumber
\eq   
(where $\d_{\G}(i) = \delta (x_i-y_i)\G^i$).\par
Let $K = k_{\mu} \cdot \G^{\mu}$. The identity 
${1\over 2} (K\cdot Z + Z \cdot K) = k_{\mu}x^{\mu}\cdot \I$ allows us 
to express the solutions to the free equations of motion 
for the bosonic massive field $\Phi$ in terms of the 
matrix-coordinate $Z$-representation. Indeed, 
if $K\cdot K = m^2\cdot \I$, the equation
\begin{eqnarray}
(D^2 {\partial_Z}^2 +m^2)\Phi &=& 0 
\end{eqnarray}
admits solutions which can be written as
\begin{eqnarray}
\relax \Phi (Z) &=& \int dV_K [ a(K) e^{{\textstyle{i\over 2}}(K\cdot Z + 
Z\cdot K)}+ a^{*}(K) e^{-{\textstyle{i\over 2}}(K\cdot Z + 
Z\cdot K)}]
\eq
where the modes $a(K)$ can be expanded in Laurent expansion as
$a(K) = \sum_{n\in \ZZ} a_n K^n$ and the $a_n$ coefficients for our
scopes can be assumed to be $c$-numbers.\par
At least for this particular case within the single-coordinate matrix
formalism we are able to recover the results obtained in the standard 
framework. The mentioned feature that the $\slash{\! \! \! \partial}$
derivative need not be associated with only spinorial fields arises as
a byproduct.

\vspace{0.2cm}
\noindent{\section{The conformal invariance in the matrix-approach.}}

The matrix nature of the coordinate in the matrix calculus 
introduces noncommutative features. In this section we
discuss this topic and show how the conformal invariance can be recovered
within such a formalism.\par
At first it should be noticed that even and odd powers of $Z$ behave
differently. Due to the previous section results we get that
$Z^{2n}= (x^2)^n\cdot\I $ is proportional to the identity, 
while $Z^{2n+1} = (x^2)^n \cdot Z$.
As a consequence only the subclass of ``odd" transformations of the kind 
$Z\mapsto Z^{2n+1}$ admits a realization in the ordinary spacetime 
coordinates $x_\mu$ as $x_\mu\mapsto x_\mu (x^2)^n$, for any 
integer-valued 
$n$. ``Even" transformations 
(i.e. mappings $Z\mapsto Z^{2n}$) cannot be realized on the
$x_\mu$ coordinates,
while they are acceptable transformations in the $Z$-coordinate 
realization.\par
Simple algebraic manipulations show that 
the {\em left} action of the $\partial_Z$
derivative on powers of $Z$ leads to
\bq
\partial_Z Z^{2n} &=& {\textstyle {2n\over D}}Z^{2n-1}\nonumber\\
\partial_Z Z^{2n+1} &=& ({\textstyle{2n +D\over D}}) Z^{2n}
\eq
The commutation relation between $Z$ and $\partial_Z$ is given by
\bq
\relax [\partial_Z , Z] 
&=& \I -{\textstyle{2\over D}} \cdot l_{\mu\nu}\Sigma^{\mu\nu}
\label{commut}
\eq
where $l_{\mu\nu}$ and $\Sigma^{\mu\nu}$ are respectively the
spacetime and spinorial generators of the Lorentz algebra:
\bq
l_{\mu\nu} &=& x_\mu\partial_\nu - x_\nu\partial_\mu\nonumber\\
\relax \Sigma^{\mu\nu} &=& {\textstyle{1\over 4}}[\G^\mu,\G^\nu]
\label{lorentzgen}
\eq
The extra-term on the r.h.s. of (\ref{commut}) is 
clearly absent in $D=1$ 
dimension. All the informations 
that we are dealing with a higher dimensional 
spacetime are therefore encoded in this extra 
operator \footnote{ It 
is tempting
to regard (\ref{commut}) as a deformation (depending on
a $\k={\textstyle {1\over D}}$ parameter) of the standard 
commutator. Perhaps $D$-dimensional
relativistic theories could therefore be analyzed in
the light of the deformation theory which, in a different context,
has been employed to recover quantization from classical structures
(see e.g. \cite{flato}). However we will not elaborate more on such  
aspects in the present paper.}.\par
The $D$-dimensional conformal 
group is defined as the set of transformations
leaving invariant the relation $ds^2=0$. It is well known that for $D>2$
the number of generators $n_C$ in the conformal group is given by
$n_C = n_L + 2D +1$, 
where $n_L = {\textstyle{1\over 2}}(D^2-D)$ is the number
of generators in the Lorentz group. The extra generators are given by the
$D$ translations, the $D$ special conformal transformations plus the 
dilatation. 
In the matrix-coordinate realization this result is recovered as
follows. While the Poincar\'e 
generators have been discussed in the previous
section and the dilatation is simply given 
by $Z\mapsto \l Z$, the only crucial
points concerns how to obtain the $D$ 
special conformal transformations. They
are given by the composition of the Poincar\'e transformations with the
conformal inversion, which in our case is expressed through the 
transformation 
\bq
Z&\mapsto& Z'= {1\over Z}
\label{inversion}
\eq
(i.e. $x_\mu\mapsto {x_\mu\over (x^2)}$). It is a simple algebraic check 
to prove that $dZ^2=0$ is preserved by (\ref{inversion}). 
No other power transformation of $Z$ for a different value of the exponent 
shares this
feature. For instance $dZ^{2n+1}\cdot dZ^{2n+1}$ is not proportional to 
$ds^2$ because an extra contribution of the kind
\begin{eqnarray} 
&& 
4n(n+1) (x^2)^{2n-1} dx_\alpha x^\a dx_\b x^\b, \nonumber
\end{eqnarray}
which vanishes only for $n=-1$, is present. This 
one and similar other consistency checks make ourselves comfortable 
with the intrinsic coherency of the matrix-coordinate formalism.\par
In the $D=1$ dimension the conformal group coincides classically with 
the $1$-dimensional 
diffeomorphisms group whose algebraic structure is given by the 
infinite-dimensional Witt algebra (centerless Virasoro algebra), spanned by
the $l_n$ generators
\bq
l_n &=& -z^{(n+1)}{\partial\over\partial z}
\eq
which satisfy the commutation relations
\bq
\relax [l_n, l_m] &=& (n-m) l_{n+m}
\label{witt}
\eq
We expect that this algebra should be recovered in higher 
dimensions as well.
This is the case indeed. If we define for any given $D$
\bq
L_n &=& -{D\over 2} Z^{2n+1}{\partial_Z}
\eq
then the $L_n$ generators satisfy 
(\ref{witt}), i.e. $[L_n,L_m] = (n-m) L_{n+m}$.
\par
In the special $D=1$ limit this algebra coincides with the Witt subalgebra
spanned by the ``even" generators ${\textstyle{1\over 2}} l_{2n}$; such
a subalgebra 
coincides with the Witt algebra itself \footnote{ It is a property of the 
Virasoro algebra that any subalgebra spanned by 
${\tilde l}_n ={\textstyle{1\over k}}l_{kn}$ generators for any given 
positive integer $k$, is still equivalent to the full Virasoro algebra. If
the non-trivial cocycle for the central extension is chosen to be of the
form $c n^3\d_{n+m,0}$, then the central charge ${\tilde c}$ present in 
the ${\tilde l}_n$ subalgebra is rescaled to be ${\tilde c} = k c$.}.
\par
If we do not limit ourselves to consider ``even" generators, but we enlarge
the structure to accomodate ``odd" generators of the kind
$M_n = - {D\over 2} Z^{2n} {\partial_Z}$ in $D>1$, 
then we no longer
find a closed algebraic structure since the commutator between $M_n$, $L_m$
involves extra operators
\bq
\relax [M_n,L_m] &=& ({\textstyle{2m-2n + 
D\over 4}})\cdot Z^{2n+2m}\cdot\partial_Z
-{\textstyle {3\over 4}} Z^{2n+2m} \cdot l_{\mu\nu}\Sigma^{\mu\nu}
\cdot \partial_Z
\eq
A closed linear algebraic structure should therefore 
include the extra operators (the second term in the r.h.s.) 
and any other new operator arising from the commutation of the      
previous ones, a procedure which has been encountered for instance
when dealing with $W_{\infty}$-algebra structures, see \cite{toppan}
and references therein.\par
Let us conclude this section by pointing out that no contradiction is 
present with the previous result that the conformal algebra in higher
dimension is finite-dimensional. Indeed only in $D=1$ the 
Witt algebra admits a geometrical interpretation as a conformal algebra.  
We have seen that for $D>2$ the conformal relation 
$dZ^2=0$ is preserved by a group of transformations with a finite number of
generators only (while the $D=2$ case can be treated within the standard 
conformal calculus). 
 
\vspace{0.2cm}
\noindent{\section{$\G$-matrices and basic notations.}}

In the two previous sections we have investigated the single
matrix-coordinate formalism and explained in some detail how
it works. To be able to go a step further and analyze the $2D$
matrix-coordinates approach we need at first to check
whether is it possible to solve the
three conditions (from ${\em i)}$ to ${\em iii)}$) formulated
in the introduction. This can be done only when properties of
$\G$-matrices for any spacetime are taken into account. For that reason 
this section
is devoted to analyze $\G$ matrices and establish our notations
and conventions. Concerning this material, we have used \cite{gamma}
as basic references.\par
A $\G$-structure associated to a given spacetime, is a matrix
representation of the Clifford algebra generators $\Gamma ^{\mu}$,
which satisfy the anticommutation relations 
\begin{equation}
\Gamma ^\mu \Gamma ^\nu +\Gamma ^\nu \Gamma ^\mu =2\eta ^{\mu \nu }%
\leavevmode\hbox{\small1\kern-3.8pt\normalsize1}_\Gamma
\label{gammaplus}
\end{equation}
(here $\eta^{\mu\nu}$ is any (pseudo)-euclidean metric in $D$ dimension).
The representation is realized 
by $2^{\textstyle{D\over 2}}\times 2^{\textstyle{D\over 2}}$
matrices which further satisfy the unitarity requirement
\begin{eqnarray}
{\G^{\mu}}^{\dagger} &=& {\G^{\mu}}^{-1}
\end{eqnarray}
as well as the tracelessness condition
\begin{eqnarray}
tr{\G^{\mu}} &=& 0
\end{eqnarray}
for any ${\mu}$.
\par
The commutator is
\begin{equation}
\Gamma ^\mu \Gamma ^\nu -\Gamma ^\nu \Gamma ^\mu 
=4\Sigma ^{\mu \nu }
\leavevmode\hbox{\small1\kern-3.8pt\normalsize1}_\Gamma
\label{gammaminus}
\end{equation}
$\S^{\mu\nu}$, already introduced in (\ref{lorentzgen}), is the generator of
the Lorentz (pseudo-rotations) group.\par
For a matter of convenience and without loss of generality 
we can work in the 
so-called Weyl representation for $\G^{\mu}$, which occours
when the dimensionality $D$ of the spacetime is even; the
$\Gamma ^{\mu}$ are block-diagonal 
\begin{eqnarray}
\Gamma ^\mu &=&\left( 
\begin{array}{cc}
0 & \sigma ^\mu \\ 
\tilde{\sigma}^\mu & 0
\end{array}
\right) 
\label{weyl}
\end{eqnarray}
The dimensionality of the $\sigma$, ${\tilde{\sigma}}$
matrices is 
$dim_{\sigma} = dim_{{\tilde{\sigma}}} 
= 2^{{\textstyle {D\over 2}}-1}$.\\
It is worth mentioning that all the results found in 
the present paper are
representation-independent and not specific of the above 
presentation.\par
Any generic $Y$ matrix, constructed with $\G$-matrices and 
their products, have spinorial 
transformation properties
(dotted and undotted indices) of the following kind
\begin{eqnarray}
Y &=&\left( 
\begin{array}{cc}
\left. {\star}_\alpha ^{}\right. ^\beta & 
{\star}_{\alpha \dot{\beta}} \\ 
{\star}^{\dot{\alpha}\beta } & \left. 
{\star}^{\dot{\alpha}}\right. _{\dot{\beta}}
\end{array}
\right) 
\label{dotted}
\end{eqnarray}
The extra matrix $\G^{D+1}$, introduced with the correct
normalization in (\ref{gammafivegen}), together with the $D$ $\G^\mu$ 
satisfy the 
(\ref{gammaplus}) and (\ref{gammaminus}) algebra in
$(D+1)$-dimensions and is block-diagonal (the blocks have equal size)
\[
\G^{D+1}=\left( 
\begin{array}{cc}
{\I}_{\s}  & 0 \\ 
 0 & - {\I}_{{\tilde\s}}
\end{array}
\right) 
\]
In terms of $\sigma^\mu$ 
and ${\tilde {\sigma}}^\mu$ the (\ref{gammaplus})
and (\ref{gammaminus})
algebra reads as follows
\begin{eqnarray}
\sigma ^\mu \tilde{\sigma}^\nu +\sigma ^\nu 
\tilde{\sigma}^\mu &=& 2\eta ^{\mu
\nu }\leavevmode\hbox{\small1\kern-3.8pt\normalsize1}_\sigma
\nonumber\\
\tilde{\sigma}^\mu \sigma ^\nu 
+\tilde{\sigma}^\nu \sigma ^\mu &=& 2\eta ^{\mu
\nu }\leavevmode\hbox{\small1\kern-3.8pt
\normalsize1}_{\tilde{\sigma}}
\label{pluscase}
\end{eqnarray}
and respectively
\begin{eqnarray}
\sigma ^\mu \tilde{\sigma}^\nu -\sigma 
^\nu \tilde{\sigma}^\mu &=& 4\sigma
^{\mu \nu }\leavevmode\hbox{\small1\kern-3.8pt\normalsize1}_\sigma
\nonumber\\
\tilde{\sigma}^\mu \sigma 
^\nu -\tilde{\sigma}^\nu \sigma ^\mu &=& 4\tilde{%
\sigma}^{\mu \nu }
\leavevmode\hbox{\small1\kern-3.8pt\normalsize1}_{\tilde{%
\sigma}}
\label{minuscase}
\end{eqnarray}
while
\[
\Sigma ^{\mu\nu} =\left( 
\begin{array}{cc}
\sigma^{\mu\nu} & 0 \\ 
0 & \tilde{\sigma}^{\mu\nu} 
\end{array}
\right) 
\]
From the even-dimensional euclidean $\G$-matrices we can 
reconstruct the 
$\G$-matrices
for any other signature by applying a Wick rotation, realized
as follows: let ${\overline \mu}$ be a direction with $-$
signature. The correponding $\G^{\overline\mu}$ is obtained
from the euclidean ${\G_E}^{\overline \mu}$ through
${\G_E}^{\overline \mu} \mapsto \G^{\overline \mu} = i
{\G_E}^{\overline \mu}$, i.e.
\begin{eqnarray}
{\sigma_E}^{\overline \mu} &\mapsto & \sigma^{\overline\mu} = 
i{\sigma_E}^{\overline\mu}\nonumber\\
{{\tilde{\sigma}}_E}^{\overline\mu} &\mapsto & 
{\tilde \sigma}^{\overline\mu} = i 
{{\tilde \sigma}_E}^{\overline\mu}
\label{wick}
\end{eqnarray}
The $\G^\nu$ matrices along the timelike $\nu$ directions 
($\eta^{\nu\nu}=+1$) are left unchanged. The absolute sign in 
the (\ref{wick}) transformations is just a matter of choice.\par
The adjoint, complex and transposed structures can be introduced 
in terms
of three unitary matrices, conventionally denoted as $A,B,C$ 
in the literature \cite{gamma}, satisfying 
\begin{eqnarray}
{\Gamma ^\mu }^{\dagger }&=&(-1)^{t+1}A\Gamma ^\mu A^{\dagger }
\\ 
{\Gamma ^\mu}^{*} &=&\eta B\Gamma ^\mu B^{\dagger } \\ 
{\Gamma ^\mu }^T &=&\eta (-1)^{t+1}C\Gamma ^\mu C^{\dagger }
\label{ABC}
\end{eqnarray}
$\eta$ is here a sign 
($\eta =\pm 1$) which in principle can be evaluated but
need not be specified for our purposes.\par 
In the euclidean (positive signature $+...+$) and only in 
the euclidean
case the $\G^\mu$ matrices can be assumed all hermitians 
(${\G^{\mu}}^{\dagger} = \G^{\mu}$ for any $\mu$). \par
For simplicity in the following the three above structures, 
adjoint, 
complex and transposed,
will also be referred as $A,B,C$-structures. 
For completeness let us report here the following properties 
satisfied by $A,B,C$: 
\begin{eqnarray}
A &=&\Gamma ^0\cdot ... \cdot \Gamma ^{t-1} \\ 
B^T &=& \varepsilon B, \\ 
C&=& B^TA
\end{eqnarray}
$\varepsilon$ is a sign ($\varepsilon =\pm 1$)
which is expressed \cite{gamma} through
$\varepsilon =\cos \frac \pi 4(s-t)-\eta \sin \frac
\pi 4(s-t)$ (as before $t$ and $s=D-t$ denote respectively
the number of timelike and spacelike coordinates). In the
formula for $A$ the product of $\G$ is restricted to timelike
coordinates only. \par
We have furthermore
\begin{eqnarray}
A^{-1}&=&(-1)^{\frac{t(t-1)}2}A \\ 
A^{*}&=&\eta ^tBAB^{-1} \\ 
A^T&=&\eta ^tCAC^{-1} \\ 
C^T&=&\varepsilon \eta ^t(-1)^{\frac{t(t-1)}2}C
\end{eqnarray}

\vspace{0.2cm}
\noindent{\section{An algorithm to iteratively compute $\G$ matrices.}}

In this section we present an algorithm which encodes fractal 
properties of the $\G$-matrices and allows to iteratively compute 
$\G$-matrices in any dimension and for any signature of the space-time
by the knowledge of lower-dimensional $\G$-matrices. As a consequence
the computation of any set of $D$-dimensional $\G$-matrices 
satisfying the (\ref{gammaplus}) algebra
is recovered from the sole knowledge of the three Pauli matrices.
\par
The algorithm here presented is central for our analysis of the $2D$
matrix-coordinates formalism in the ``splitting" case and is also
quite useful in proving the vector-contraction 
identities we introduce and discuss
in the next section.\par
The $\G$-matrices in even $D$ spacetime dimension can be represented
from the $\g$-matrices in $(p+1)$ and $(q+1)$ spacetime dimensions 
(we will
use capital and lower letters for reasons of typographical clarity) 
where the even integers $p,q$ satisfy the condition 
\begin{eqnarray}
D&=& p+q+2
\label{dimensionality}
\end{eqnarray}
Since, as recalled in the previous section, $\G$-matrices for any
signature are obtained from the euclidean $\G$-matrices through 
a Wick
rotation, it is sufficient to present our formulas in the case when
all the $\G$-matrices involved (in $(p+1)$, $(q+1)$ and $D$ dimensions)
are euclidean.\par
The capital index $M=0,1,..., D-1$ is used to span the 
$D$-dimensional space, while $m=0,1,...,p$ and ${\overline m} =
0,1,...,q$ are respectively employed for the $(p+1)$ and 
$(q+1)$-dimensional spaces. The corresponding $\G$-matrices
will be denoted as ${\G_D}^M$, ${\g_p}^m$, ${\g_q}^{\overline m}$.
\par
The symbol $\I_n$ will denote the $2^{\textstyle{n\over 2}}\times
2^{\textstyle{n\over 2}}$ identity matrix.  \par
For concision of notations the symbols $\I_0$ and ${\g_0}^0$ (i.e.
the ``$1$-dimensional $\G$-matrix") will both denote the constant 
number $1$.\par
It is a simple algebraic exercise to prove that the set of ${\G_D}^M$
matrices can be realized through the position
\begin{eqnarray}
{\G_D}^{M}&=&\left( 
\begin{array}{cc}
0  & \I_q\otimes {\g_p}^{m};\quad -i {\g_q}^{\overline m}\otimes \I_p \\ 
\I_q\otimes {\g_p}^{m};\quad i {\g_q}^{\overline m}\otimes \I_p  
& 0
\end{array}
\right) 
\label{gammas}
\end{eqnarray}
with $M\equiv (m, p+1+{\overline m})$.\par
The condition (\ref{dimensionality}) is necessary in order to match
the dimensionality of the $\G$-matrices in the left and right side
(due to (\ref{dimensionality}) the dimension of the r.h.s. matrix is 
$2\cdot
2^{\textstyle{p\over 2}}\cdot 
2^{\textstyle {q\over 2}}=2^{\textstyle{D\over 2}}$ if 
(\ref{dimensionality}) is taken into account).   
A further consequence of the (\ref{dimensionality}) condition is that the 
``generalized $\g^5$-matrices" (\ref{gammafivegen}) of the kind ${\g_p}^p$
and ${\g_q}^q$ are necessarily present, which implies a decomposition
of the even dimensional $D$ spacetime into two {\em odd}-dimensional
$p+1$ and $q+1$-spacetimes.\par
The decomposition realized by (\ref{gammas}) works for any couple of even 
integers $p,q$ satisfying the (\ref{dimensionality}) condition. This 
implies that
for any given even integer $D$ the number $n_D$ of inequivalent 
decompositions 
(factoring out the ones trivially obtained by exchanging 
$p\leftrightarrow q$) is given by
$n_D = {\textstyle{1\over 4}} (D+ r)$, where either $r=0$ or $r=2$ 
according respectively if $D$ is a multiple of $4$ or not. At the lowest 
dimensions we have the following list of allowed decompositions:
\begin{eqnarray}
D=2 & \leftarrow & \{(p=0, q=0)\}\nonumber\\
D=4 & \leftarrow & \{(p=0, q=2)\}\nonumber\\
D=6 & \leftarrow & \{(p=0, q=4), \quad (p=2, q=2)\}\nonumber\\
D=8 & \leftarrow & \{(p=0, q=6), \quad (p=2, q=4)\}\nonumber\\
D=10 & \leftarrow & \{(p=0, q=8), \quad (p=2, q=6), \quad (p=4, q=4)\}
\end{eqnarray}
and so on. It is worth mentioning here that in issues involving
Kaluza-Klein compactifications to lower-dimensional spacetimes the above
result can find useful applications in suggesting which one of the
allowed decompositions is the most convenient to choose.
\par
Since the formula (\ref{gammas}) is quite important for our purposes
it is convenient to furnish it in two other presentation. We can
write in the Weyl realization
\begin{eqnarray}
{\sigma_D}^M &=& 
(\I_q\otimes {\g_p}^{m};\quad -i {\g_q}^{\overline m}\otimes \I_p) 
\nonumber\\
{{\tilde\sigma}_D}^M &=& 
(\I_q\otimes {\g_p}^{m};\quad i {\g_q}^{\overline m}\otimes \I_p)
\end{eqnarray}
${\G_D}^M$ can also be expressed through
\begin{eqnarray}
{\G_D}^m &=& \tau_x \otimes \I_q\otimes {\g_p}^m\nonumber\\
{\G_D}^{p+1+{\overline m}} &=& \tau_y \otimes {\g_q}^{\overline m}
\otimes \I_p
\end{eqnarray}
with the help of the off-diagonal Pauli matrices $\tau_x$, $\tau_y$.\par
The three Pauli matrices given by 
\begin{equation}
\tau _x=\left( 
\begin{array}{cc}
0 & 1 \\ 
1 & 0
\end{array}
\right) ,\quad \tau _y=\left( 
\begin{array}{cc}
0 & -i \\ 
i & 0
\end{array}
\right) ,\quad \tau _z=\left( 
\begin{array}{cc}
1 & 0 \\ 
0 & -1
\end{array}
\right) \label{pauli}
\end{equation}
can be regarded as the $\G$-matrices for the euclidean
three-dimensional space. It is evident that any 
$D$-dimensional $\G$-matrix can be constructed, with 
repeated applications of the (\ref{gammas}) formula,
by tensoring the (\ref{pauli}) Pauli matrices. The
statement made at the beginning of this section is
therefore proven.\par
Let us make some comments about the algebraic meaning of the formula 
(\ref{gammas}). The generators $\S^{MN}$ of the Lorentz transformations
are expressed through the commutators of $\G$ matrices, see 
(\ref{lorentzgen}).
By using the (\ref{gammas}) decomposition the index $M$ is splitted into
$m$, ${\overline m}$ indices. The Lorentz algebra ${\cal G}$ admits a
decomposition in three subspaces ${\cal M}_+$, ${\cal M}_-$
and ${\cal K}$, spanned respectively by
the generators $\S^{mn}\in {\cal M}_+$, 
$\S^{{\overline m}{\overline n}}\in {\cal M}_-$
and $\S^{m {\overline n}}\in {\cal K}$. \par
${\cal M}_+$ and ${\cal M}_-$ are the ${\cal G}$ subalgebras corresponding
to the Lorentz algebras for the $(p+1)$-dimensional and 
respectively the $(q+1)$-dimensional subspaces entering the 
(\ref{gammas}) decomposition. By setting 
${\cal M} =_{def} {\cal M}_+ \oplus
{\cal M}_-$, the full Lorentz algebra is expressed as 
\begin{eqnarray}
{\cal G} &=& {\cal M} \oplus {\cal K}
\label{decomp}
\end{eqnarray}
The Lorentz commutators in ${\cal G}$ satisfy the following set of
symbolic relations
\begin{eqnarray} 
\relax [ {\cal M}, {\cal M} ] &=& {\cal M}\nonumber\\
\relax [ {\cal M}, {\cal K} ] &=& {\cal K}\nonumber\\
\relax [ {\cal K}, {\cal K} ] &=& {\cal M}
\end{eqnarray}
The existence of such relations gives to the Lorentz algebra ${\cal G}$
the structure of a homogeneous space w.r.t. its (\ref{decomp})
decomposition. The existence of the (\ref{gammas}) representation for the
$\G$-matrices is just a reflection of such a homogeneity property.

\vspace{0.2cm}
\noindent{\section{The $2D$ matrix formalism in the ``vector-trace" 
approach.}}

In this section we start discussing how to implement our
program which prescribes the introduction of two distinct 
$Z$, ${\overline Z}$ matrix-coordinates. We recall that the
basic properties required (from ${\em i)}$ to ${\em iii b)}$ )
have already been presented in the introduction. \par
It is quite evident that we have no longer the possibility to
identify one of the coordinates (let's say $Z$) with the position
$Z=x_\mu
\G^{\mu}$ as in the single matrix-coordinate formalism, since in this 
case no room is left to introduce the second coordinate ${\overline Z}$, 
commuting with the previous one and satisfying 
$d{\overline Z}\cdot dZ = ds^2\cdot\I$. 
A different strategy has to be employed. In this section we present one,
which we conventionally call the ``trace" approach since, as we will see, 
it involves some identities concerning contractions of vector indices
of $\G$-matrices (``vector traces"). Another approach based on
a different construction will be discussed in the next sections. 
An important feature which should be stressed
here is the fact that the requirements put by the $2D$ matrix-coordinates
formalism lead to some non-trivial constraints concerning the structure
of space-times. Different matrix-solutions can be found to our program 
depending on the dimensionality and the signature of the spacetimes.\par
We gain much more freedom to investigate our problem if we take as 
starting building blocks to construct
matrix-valued objects not just the $\G^\mu$-matrices themselves, 
but instead the $\s^{\mu}$, ${\tilde\s}^{\mu}$ blocks 
(together with their conjugated matrices under adjoint, transposed or 
complex
action) 
entering the Weyl realization (\ref{weyl}).\par
Let us introduce in order to simplify notations 
\begin{eqnarray}
&& \omega = x_\mu\s^\mu, \quad \quad
{\tilde\omega} = x_\mu {\tilde\sigma}^\mu
\end{eqnarray}
The spinorial (dotted and undotted indices) 
transformation properties for $\omega$, ${\tilde{\omega}}$
and their $A,B,C$-transformed quantities 
are as follows
\begin{eqnarray}
&&
\{ \omega, {\tilde \omega}^\dagger, \omega^*, {\tilde\omega^T}\}
 \equiv  \star_{\a{\dot\b}},\quad\quad
\{ {\tilde\omega}, {\omega^\dagger}, {\tilde{\omega}}^*, \omega^T\}
\equiv  \star^{{\dot\alpha} \beta}
\label{omegas}
\end{eqnarray}
In accordance with the above transformation properties the first
and the second set of matrix-valued objects have to be inserted in
matrices of the kind of (\ref{dotted}) respectively in the upper right
(lower left) corner. \par
The $\s$'s and ${\tilde \s}$'s matrices satisfy the anticommutation 
and commutation relations given by (\ref{pluscase}) and
(\ref{minuscase}). Analogous relations are 
immediately obtained by applying on them the $A,B,C$-transformations
(\ref{ABC}). 
The requirement of commutativity 
($\relax [ Z, {\overline Z}]=0$), as well as
the disentangling of the coordinates under the {\em left} action of 
derivatives
(i.e. $\partial_Z {\overline Z} = \partial_{\overline Z} Z =0 $) can be 
solved
with the help of the (\ref{pluscase}) relations. They apply however only if
at most a single matrix of the kind of $\omega$, ${\tilde \omega}$ 
(or their conjugated quantities) is inserted in the upper right or lower 
left
diagonal block of a bigger matrix (\ref{dotted}) to build up
$Z$, ${\overline Z}$. For that reason we 
do not consider here the possibility that mixed terms could be
present. The investigation about the possibility to solve the above 
relations
in this context is much more involved and does not seem to use general
arguments as the case we are analyzing here. It is therefore left as an
open problem for further investigations.
On the other hand the construction involving single blocks is here fully 
analyzed and the complete solution is furnished.\par
We ask for matrix-valued $Z$, ${\overline Z}$ of the kind
\begin{equation}
Z=\left( 
\begin{array}{cc}
0 & \omega\\ 
\star & 0
\end{array}
\right) ,\quad 
{\overline Z} =\left( 
\begin{array}{cc}
0 & \star\\ 
{\tilde \omega} & 0
\end{array}
\right) \label{zz} 
\end{equation}
To keep covariance the $\star$ in the above formulas should be 
replaced either by the 
$0$-matrix 
or by the matrices in (\ref{omegas}) with the right covariance
properties. The commutation requirement $[Z, {\overline Z}] =0$ rules
out the possibility to use the $0$-matrix so that the only left
possibilities are either
\begin{equation}
Z=\left( 
\begin{array}{cc}
0 & \omega\\ 
\omega^\# & 0
\end{array}
\right) ,\quad 
{\overline Z} =\left( 
\begin{array}{cc}
0 & {\tilde\omega}^\#\\ 
{\tilde \omega} & 0
\end{array}
\right)  \label{zdiesis}
\end{equation}
 or
\begin{equation}
Z=\left( 
\begin{array}{cc}
0 & \omega\\ 
{\tilde\omega}^* & 0
\end{array}
\right) ,\quad 
{\overline Z} =\left( 
\begin{array}{cc}
0 & {\omega}^*\\ 
{\tilde \omega} & 0
\end{array}
\right) \label{zstar} 
\end{equation}
(since the transposed and the adjoint case are formally similar it is
convenient to introduce a unique symbol $\#$ to denote both of them,
i.e. $\#\equiv T, \dagger$).\par
In both the above cases the identification of $Z, {\overline Z}$ through
either (\ref{zdiesis}) or (\ref{zstar}) implies that 
the commutativity property is satisfied in consequence 
of (\ref{pluscase}). In the two $\#$-cases above 
the commutativity requires for instance the vanishing of the expression
\begin{eqnarray}
&& x_\mu x_\nu ( \s^\mu{{\tilde\s}^\nu}- {{\tilde\s}^{\nu\;\#}}
{\s^\mu}^\# )
\end{eqnarray}
This is indeed so as it can be realized by expanding the term inside
the parenthesis in its symmetric and antisymmetric component 
under the $\mu\leftrightarrow \nu$ exchange. 
Notice the role of the $-$ sign and the fact that $Z$, ${\overline Z}$ in
(\ref{zdiesis}) are correctly ``fine-tuned" in order to guarantee the
commutativity. A similar analysis works for the complex $*$-case as well.
\par
The (\ref{pluscase}) identities imply the following relations
\begin{eqnarray}
&& \sigma^\mu {\tilde \sigma}_{\mu} = D\cdot \I_\s
\quad\quad {\tilde\s}^\mu\s_\mu = D\cdot \I_{\tilde\s}
\label{dimension}
\end{eqnarray}
(where from now on the Einstein convention over repeated indices is 
understood).\par
Such identities allow us to introduce the derivative $\partial_Z$, 
$\partial_{\overline Z}$ which satisfy the ${\em ii)}$ condition and the 
normalization requirement ${\em iii b)}$. They are given in the
$\#$-cases by
\begin{equation}
\partial_Z= {{1\over D}}\left( 
\begin{array}{cc}
0 & \partial_\mu {{\tilde\s}^{\mu\;\#}}\\ 
\partial_\mu {{\tilde \s}^{\mu}} & 0
\end{array}
\right) ,\quad 
\partial_{\overline Z} ={1\over D}\left( 
\begin{array}{cc}
0 & \partial_\mu {\sigma^{\mu}}\\ 
\partial_\mu{{\s^{\mu\; \#}}} & 0
\end{array}
\right)  \label{diesisder}
\end{equation}
and in the $*$-case by
\begin{equation}
\partial_Z= {{1\over D}}\left( 
\begin{array}{cc}
0 & \partial_\mu {{\s}^{\mu\; *}}\\ 
\partial_\mu {{\tilde\s}^\mu} & 0
\end{array}
\right) ,\quad 
\partial_{\overline Z} ={1\over D}\left( 
\begin{array}{cc}
0 & \partial_\mu{\sigma^\mu}\\ 
\partial_\mu{{{\tilde \s}^{\mu\;*}}} & 0
\end{array}
\right) \label{complexder}
\end{equation}
In both the $\#$ and $*$-cases we have the relation
\begin{eqnarray}
&& \partial_Z\partial_{\overline Z} =\partial_{\overline Z}\partial_Z=
{{1\over D^2}}\Box \cdot \I
\end{eqnarray}
Up to now all the properties required by the $2D$ matrix-coordinates
formalism have been satisfied. The last property which should be 
implemented,
but a fundamental one, is the ``disentangling condition" 
${\em iii a)}$.\par
One can immediately check that 
$\partial_Z {\overline Z} =\partial_{\overline Z} Z = 0$
is satisfied whether, according to the different cases, the following 
contractions of the vector indices give
vanishing results:
\begin{eqnarray}
&&
{\cal A} \equiv  \s^\mu {\s_\mu}^\dagger, \quad
{\cal B} \equiv  \s^\mu {{\tilde\s}_\mu}^*, \quad
{\cal C} \equiv  \s^\mu {{\s}_\mu}^T 
\end{eqnarray}
(and similarly
${\tilde {\cal A}} = {\tilde \s}^\mu {{\tilde \s}_\mu}^\dagger$,
${\tilde{\cal B}} = {{\tilde\s}^{\mu\; *}}{\s_\mu}$ and 
${\tilde{\cal C}} = {{\tilde \s}^\mu}{{\tilde\s}_\mu}^T$ should 
vanish as well).
${\cal A}$, ${\cal B}$, ${\cal C}$ 
are all proportional to ${\I}_\s$ 
with a
proportionality factor $a$, $b$, $c$ respectively (one can easily realize
that ${\tilde{\cal A}}$, ${\tilde{\cal B}}$, ${\tilde{\cal C}}$ are 
proportional to ${\I}_{\tilde{\s}}$ with the same $a$, $b$, $c$ constant 
factors).\par
An equivalent way of rephrasing the above properties reads as follows
\begin{eqnarray}
\G^\mu{\G_\mu}^\dagger &=& a\cdot {\I}_\G\nonumber\\
\quad \G^\mu{\G_\mu}^* &=& b\cdot {\I}_\G\nonumber\\
\G^\mu{\G_\mu}^T &=& c\cdot{\I}_\G
\label{vectortraces}
\end{eqnarray}
We are therefore left to determine under which conditions the above 
$a$, $b$, $c$ constants are vanishing.\par
Before going ahead let us however point out that while $a$ is 
always representation-independent, $c$ in principle could not
be representation-independent (in effect it is so and is always vanishing)
and $b$ is representation-independent only in the euclidean case 
(for generic
signatures its value depend on the way the Wick rotation (\ref{wick}) 
has been
performed). The algebraic meaning of $b$ as an index labelling classes of
equivalence of $\G$-structures under conjugations determined by
{\em both} real and orthogonal matrices will be 
discussed in section $8$.\par
The remark concerning the representation-independence can be 
immediately understood when realizing that a different $\G$-structure
satisfying the Weyl condition is recovered from the $\G^\mu$ by
by
simultaneously rescaling all $\s$'s and ${\tilde\s}$'s through 
\begin{eqnarray}
\s^\mu &\mapsto & -i\s^\mu\nonumber\\
{\tilde\s}^\mu &\mapsto & i{\tilde\s}^\mu
\label{flips}
\end{eqnarray}
Under such a transformation $a$, $b$, and $c$ are mapped 
as follows: $a\mapsto a$, $b\mapsto -b$, $c\mapsto -c$.\par
The above transformation can also be reexpressed with the help of
the Pauli matrices as
\begin{eqnarray}
&& \G^\mu       
\mapsto (i \tau_z\otimes \I )\cdot
\G^\mu 
\end{eqnarray}
In the next section we compute the coefficients
$a$, $b$, $c$ for any even-dimensional spacetime.

\vspace{0.2cm}
\noindent{\section{The vector-contraction identities.}} 

In the previous section we have furnished the motivations why we are 
interested in computing the ``vector-contractions" expressed 
by the formula (\ref{vectortraces}),
i.e. the coefficients $a$, $b$, $c$. Here we furnish the results 
together with their proofs.
\par
The following properties hold:
\begin{eqnarray}
i)\quad a &=& t-s
\label{avalues}
\end{eqnarray}
where, as usual, $t$ ($s$) denotes the number of timelike (spacelike)
directions in $D=t+s$ dimensions; 
\begin{eqnarray}
ii) \quad b &=& 2(t_+-t_-)
\label{bvalues}
\end{eqnarray}
where $t_+$ (respectively $t_-$) are non-negative integers denoting the 
number of 
time-directions (whose total number is $t=t_++t_-$) associated
to $\G$-matrices which are symmetric (respectively antisymmetric) under
transposed conjugation in the Weyl representation; 
\begin{eqnarray}
iii) \quad c &=& 0
\label{cvalues}
\end{eqnarray}
identically in any spacetime.\par
As a result the $2D$ matrix-formalism defined in terms of  
the $A$-structure works only in $(t=n,s=n)$ spacetimes, while in terms of
the $C$-structure it is always defined for any even-dimensional spacetime.
For what concerns $b$, it can assume among other possible values, 
the $0$-value only when the spacetime admits an even number of time 
directions, under the assumption $t_+=t_-=m$, $t=2m$.\par
The $B$-structure turns out to be
defined only for spacetimes with even number of timelike (+ signature) 
and even 
number of
spacelike (- signature) directions.\par
The strategy to prove the above statements is the following. \par
For what concerns
the computation of $a$ we can start with the euclidean case. In this case 
we can 
consistently assume
\begin{eqnarray}
{\G^\mu}^\dagger &=& \G^\mu
\end{eqnarray} 
By using (\ref{pluscase}) we get that
$a$ in the euclidean is $a=D$. 
The (\ref{wick}) ``Wick rotation" applied to the 
${\overline\mu}$ direction leads to a change of sign in the contribution 
of $\G^{\overline\mu}\G_{\overline\mu}$ (indeed $+1\mapsto -1$), 
which proves the above result.\par
For what concerns $c$ we proceed as follows. At first we notice that 
any Wick
rotation leaves unchanged the contribution of the corresponding 
direction so that $c$ does not depend on the signature of the spacetime. 
It
is therefore enough to compute $c$ in the euclidean case. We can do it 
iteratively by determining the value $c_{D+2}$ of $c$ in 
($D+2$)-dimensions from its $D$-dimensional value $c_D$. 
It is convenient to do so with the help of the (\ref{gammas}) formula, 
taken with the ``extremal" decomposition $q=0$, $p=D$.
We get for ${\s_{D+2}}^{\mu}$:
\begin{eqnarray}
{\s_{D+2}}^{\mu} &=& ({\G_D}^{{\tilde\mu}};\quad 
\G^{D+1};\quad -i\cdot \I_D)
\end{eqnarray}
where ${\tilde{\mu}}=0,1,..., D-1$. The two last terms on the right 
hand side
give opposite contributions which cancel each other to the computation 
of $c_{D+2}$. Therefore $c_{D+2}=c_D$. On the other hand  
an immediate computation shows that in $D=2$, $c_2=0$. The (\ref{cvalues}) 
formula is therefore 
proven.\par
The above result is just one way of proving the well-known
property that in the Weyl representation the $\G$-matrices can all be
chosen simultaneously 
either symmetric or antisymmetric under transposition
\begin{eqnarray}
&& {\G^\mu}^T = {\e_\mu}^T \G^\mu ; \quad\quad {\e_\mu}^T =\pm 1
\label{transposed}
\end{eqnarray}
and that the number of ``symmetric" ${\mu}_+$ directions 
(${\e_{\mu_+}}^T = +1$) 
is equal to the number of ``antisymmetric" ${\mu}_-$ directions 
(${\e_{\mu_-}}^T =-1$).
\par
Due to the transformation (\ref{flips}) the (anti-)symmetric 
character of the $\G^{\overline\mu}$ along the ${\overline \mu}$ direction
(and therefore the sign in (\ref{transposed}))
is arbitrary and conventional since (\ref{flips}) reverts the
symmetry properties under transposition. The relative sign between two
arbitrary directions however is left unchanged and acquires an absolute
meaning.\par
For later purposes it is convenient to introduce the sign
${\e_\mu}^*$ as
\begin{eqnarray}
\G^{\mu\; *} &=& {\e_{\mu}}^* \G^\mu
\label{epssign}
\end{eqnarray}
We remark that ${\e_\mu}^*$ changes sign when a Wick rotation is performed
along $\mu$. \par
It should be noticed that in an even $D=t+s$ spacetime 
with $(t,s)$-signature the choice of which $\G$ matrices should be assumed 
$T$-symmetric ($T$-antisymmetric)
can be made in different ways. Let us denote with $t_+$ ($t_-$) the
number of time-directions associated to $T$-symmetric ($T$-antisymmetric) 
$\G$ matrices; $s_+$ ($s_-$) will denote the number of 
$T$-symmetric ($T$-antisymmetric) spatial directions. Clearly, 
from the previously stated results
\begin{eqnarray}
&& t= t_+ + t_-\nonumber\\
&& s= s_+ + s_-\nonumber\\
&& t_+ + s_+ = t_- + s_-= {1\over 2}(t+s)
\label{ttss}
\end{eqnarray}
It turns out that $b$ can be recovered from the computations already
performed by setting
\begin{eqnarray}
&&
\G^\mu\cdot{\G_\mu}^* = \G^\mu \cdot({\G_\mu}^T)^\dagger=\nonumber\\
&&=\G^{\mu_+}\cdot{\G_{\mu_+}}^\dagger - \G^{\mu_-}
\cdot{\G_{\mu_-}}^\dagger
= (t_+ -s_+ -t_- +s_-)\cdot \I_\G ,
\nonumber
\end{eqnarray}
that is, due to (\ref{ttss}), $b= 2(t_+-t_-)$. QED.\par
 
\vspace{0.2cm}
\noindent{\section{An index labelling the inequivalent $\G$-structures
under real orthogonal conjugation and their associated Wick rotations.}}

At this point of our analysis it is convenient to make a little digression
in order to explain the algebraic 
significance of the coefficient $b$ which can 
be reintroduced through the position ($b=2I$):
\begin{eqnarray}
I &=&  
{1\over 2^{ {D\over 2} +1}}\cdot tr(\G^\mu\cdot{\G_\mu}^*) = (t_+-t_-)
\label{index}
\end{eqnarray}
$I$ is an index with a deep algebraic meaning. We recall at first 
a fundamental
property of the $\G$-structures (defined in section $4$), known
as the ``fundamental Pauli theorem" (see \cite{sakurai}), stating 
that they are all
unitarily equivalent. This implies that given two $\G$-structures,
denoted as ${\G_I}^{\mu}$, ${\G_{II}}^{\mu}$, a  
unitary matrix $S$ ($S^{-1} = S^{\dagger}$) can always be found such that 
${\G_{II}}^{\mu} = S{\G_I}^{\mu} S^\dagger $ for any $\mu$. Moreover,
up to a normalization factor, $S$ is uniquely determined.
As a consequence the representation-independence of the Dirac equation is 
guaranteed. \par
The index $I$, as shown by the previous 
section computations,
is not defined on the above equivalence class. However one can
easily realize that
$I$ is well-defined on the class of equivalence of $\G$-structures which
are conjugated under a real and orthogonal transformation,
i.e. such that 
${\G_{II}}^{\mu} = O{\G_I}^{\mu} O^T $ for any $\mu$, with $O$ 
a real-valued matrix belonging to the orthogonal group $O(2^{D\over2})$,
$2^{D\over 2}$ being the dimensionality of $\G^{\mu}$. 
The index $I$ is therefore mathematically meaningful and can find useful
applications in issues where 
reality conditions, not just unitary equivalence,
have to be imposed. We already know that there exists spacetimes for which
$I$ assumes different values. Such spacetimes support inequivalent
$\G$-structures under real and orthogonal conjugation. 
The fact that inequivalent real structures 
arise out of a single ``complex" structure is of course not at all
surprising. In a related area we have the example of the real forms
which are associated to a given complex Lie algebra.  \par 
The index $I$ admits another interpretation. It classifies the 
inequivalent
ways a Wick rotation can be performed from the euclidean $D$-dimensional 
space to
a given ($t, s=D-t$) pseudoeuclidean space. 
We will briefly discuss this topic in the following. Our considerations
will be simplified if we analyze not just the index $I$ itself, but its 
modulus
$|I|$. By taking into 
account the (\ref{flips}) transformation, $|I|$ classifies
the equivalence-classes of $\G$-structures under conjugation for 
the tensor group
$O(2^{D\over 2})\otimes {\bf Z}_2$.\par
Without 
loss of generality (to recover the condition below it is 
sufficient to perform a $t\leftrightarrow s$ exchange), we can 
further restrict 
$t$ to be
$t \leq {D\over 2}$. 
Under this restriction the index $|I|$ for an odd number 
of time-directions ($t=2k+1$) assumes
all the possible $k+1$ different odd-values $|I|=1,3,...,2k+1$
(i.e. for $t_+=0, 1,...,k$ in the reverse order), while for an
even number of time directions ($t=2k$) it assumes all the possible 
$k+1$
even values $|I|=0,2,..., 2k$ (here again
for $t_+=0,1,...,k$ in the reverse order).  \par
Please notice that not only in the euclidean, but even in the 
generalized Minkowski case $(t=1, s= D-1)$, $|I|$ detects just one
class of equivalence. \par
In practice Wick rotations corresponding to a given value
of $|I|$ can be quite easily constructed. Let us
start with the euclidean $D=2n$ space. The directions are splitted into
two classes on $n$ elements each, according to the (anti-)symmetry
property under transposition for their associated $\G$-matrices 
(or equivalently, their ${\e_\mu}^*$  (\ref{epssign}) sign).
We can list them as $\relax[SS...S|AA...A]$ or as
$\relax [++ ...+| --...-]$. We recall that the Wick rotation flips 
the ${\e_\mu}^*$ sign so that
\begin{eqnarray}
\relax ([SS...|AA...]\equiv [++...|--...]) &\mapsto ([(iS)S...|A...]\equiv
[-+...|--...])
\end{eqnarray}
with a clear use of the symbols.\par
It is evident that for any $n$ the passage from the euclidean 
$(2n,0)$ space
to the $(1,2n-1)$-Minkowski spacetime
can be done unambiguously when (\ref{flips}) is taken into account,
indeed $|I|$ can only be $|I|=1$. However, starting from the
$t=2$ case, the Wick rotation can be done in inequivalent ways.
For instance the passage from the euclidean $(4,0)$
space to the $(2,2)$ spacetime can be done through either 
\begin{eqnarray}
i) \quad \relax [++|--] &\mapsto & [++|++]
\label{22no}
\end{eqnarray}
(i.e. $t_+=0$, $t_-=2$) so that $|I|=2$, or 
\begin{eqnarray}
ii) \quad \relax [++|--] &\mapsto & [+-|-+]
\label{22yes}
\end{eqnarray}
(i.e. $t_+=t_-=1$) with $|I|=0$.\par
Similarly, the passage $(6,0)\rightarrow (2,4)$ can be done
through either
\begin{eqnarray}
i) \quad \relax [+++|---] &\mapsto & [+++|-++]
\end{eqnarray}
($t_+=0$, $t_-=2$, $|I|=2$) or
\begin{eqnarray}
ii) \quad \relax [+++|---] &\mapsto & [++-|--+]
\end{eqnarray}
($t_+=t_-=1$, $I=0$).\par
As from the Wick rotations $(6,0)\rightarrow (3,3)$, we can have either
\begin{eqnarray}
i) \quad \relax [+++|---] &\mapsto & [+++|+++]
\end{eqnarray}
($t_+=0$, $t_-=3$, $|I|=3$), or
\begin{eqnarray}
ii) \quad \relax [+++|---] &\mapsto & [++-|-++]
\end{eqnarray}
($t_+=1$, $t_-=2$, $|I|=1$).\par
The iteration of the procedure to more general cases is now evident.\par
In all the above transformations we have picked up a Wick rotation which
is representative of its class of equivalence. 
The fact that inequivalent $\G$-structures, labelled by the index $|I|$, 
can be
associated to a given space-time has immediate consequences to our problem
of finding a $2D$ matrix-valued complex structure. Indeed, as discussed in
section $6$, the only structure which endorses a complex structure for 
the $Z$, ${\overline Z}$ matrix-coordinates is the $B$-structure. Formula
(\ref{zstar}) applies and we get
\begin{eqnarray}
{\overline Z} &=& Z^*
\end{eqnarray}
As remarked in the previous section the only spacetimes which allow a
$2D$-matrix valued complex calculus are those for which $b\equiv I =0$.
We already noticed that this implies an even number of time
coordinates (and an even number of space coordinates due to the assumption
that $D$ is even). The discussion of this section shows however 
that in
order to get a $2D$-matrix valued complex calculus, it is not sufficient
just to pick up a $(2k, 2n-2k)$ spacetime. A ``correct" Wick rotation from
the euclidean (one of those leading to $t_+=t_-=k$) has to be performed. 
For even times there is a $\G$ structure which satisfies $|I|=0$. 
Such a $\G$ structure (with its associated Wick rotations) has to be 
carefully determined. In the $(2,2)$ case for instance 
it corresponds to the
formula (\ref{22yes}), while the Wick-rotation (\ref{22no}), 
belonging to a different $\G$-structure, must be discarded. \par
We conclude this section by remarkig that issues involving two-times
physics are at present quite investigated, see e.g. \cite{bars}. 

\vspace{0.2cm}
\noindent{\section{The $2D$-matrix formalism revisited.}}

In this section we collect all the results previously obtained
concerning the $2D$-matrix calculus and present them in a single
unifying framework which makes formally similar the analysis
of the three $A$, $B$, 
$C$ cases discussed so far. The ``splitting case" $S$, whose discussion is
postponed to a later section, also fits the following formulas.\par
Let us introduce at the first the matrices 
${\O^\mu}_{(\star )}\equiv \O^{\mu}$, 
${{\overline\O}^\mu}_{(\star )}\equiv {\overline\O}^{\mu}$, 
where the $(\star)$ index denotes
one of the constructions ($A$, $B$, $C$ or $S$) which proves to work. In the
following the $(\star)$ index will be omitted in order not to 
burden the notation, 
but it should be understood.\par
The $\O$'s and ${\overline\O}$'s matrices, with 
${\overline \O}^{\mu}\neq \Omega^{\mu}$, are constructed to satisfy
the anticommutation relations
\begin{eqnarray}
\O^\mu {\overline \O}^\nu +\O^\nu{\overline\O}^\mu &=& 2\eta^{\mu\nu}\I
\nonumber\\
{\overline\O}^\mu\O^\nu +{\overline\O}^\nu\O^\mu &=& 2\eta^{\mu\nu}\I
\label{oobar}
\end{eqnarray}
An useful identity which immediately follows is
\begin{eqnarray}
&&
{\overline\O}^\mu\O_\mu =\O^\mu{\overline\O}_\mu = D\cdot\I
\end{eqnarray}
A further requirement which has been imposed is expressed by the formula
\begin{eqnarray}
&& \O^\mu {\O}_\mu ={\overline\O}^\mu{\overline\O}_\mu =0
\end{eqnarray}
(here and above the Einstein convention is understood). The latter relation,
in the $A$, $B$, $C$ cases, is a consequence of the vector-contractions 
properties of $\G$-matrices, and the conditions when is satisfied 
have been discussed
section $7$.\par
We can introduce the matrix coordinates $Z$, ${\overline Z}$, and their
relative matrix derivatives $\partial_Z$, $\partial_{\overline Z}$ through
\begin{eqnarray}
Z &=& x_\mu \O^\mu\nonumber\\
{\overline Z} &=& x_\mu {\overline \O}^\mu
\end{eqnarray}
\begin{eqnarray}
\partial_Z &=& {1\over D} \partial_\mu {\overline\O}^\mu\nonumber\\
\partial_{\overline Z} &=& {1\over D} \partial_\mu \O^\mu
\end{eqnarray}
It is convenient to formally define the following (anti)-commutators
\begin{eqnarray}
\O^\mu\O^\nu \pm \O^\nu\O^\mu &=& {\Xi_\pm}^{\mu\nu}\nonumber\\
{\overline\O}^\mu{\overline\O}^\nu \pm {\overline\O}^\nu{\overline\O}^\mu
&=& {{\overline\Xi}_\pm}^{\mu\nu}
\nonumber\\
\O^\mu{\overline\O}^{\nu}-\O^{\nu}{\overline\O}^\mu &=& \O^{\mu\nu}
\nonumber\\
{\overline\O}^\mu{\O}^\nu -{\overline\O}^\nu\O^\mu &=& 
{\overline \O}^{\mu\nu}
\end{eqnarray}
For our purposes we do not need to compute them esplicitly, however
the formulas for $\O^{\mu\nu}$, ${\overline\O}^{\mu\nu}$ will be
presented at the end.\par
The following commutation relations hold
\begin{eqnarray}
\relax [ Z, {\overline Z}] &=& 0\nonumber\\
\relax [ \partial_Z, Z] &=& [\partial_{\overline Z}, {\overline Z}] =
\I - {1\over 4D} l_{\mu\nu}({\overline\O}^{\mu\nu} + \O^{\mu\nu})
\nonumber\\
\relax [\partial_Z, {\overline Z}] &=& - {1\over 2D} l_{\mu\nu} 
{{\overline\Xi}_-}^{\mu\nu}\nonumber\\
\relax [\partial_{\overline Z} , Z ] 
&=& -{1\over 2D} l_{\mu\nu} {\Xi_-}^{\mu\nu}
\end{eqnarray}
where $l_{\mu\nu} = x_\mu\partial_\nu -x_\nu\partial_\mu$.\par
When either the $A$ ($\equiv \dagger$) or the $C$ structure ($\equiv T$) 
are
employed we have (as before $\# \equiv \dagger, T$)
\begin{eqnarray}
&& {\O_{(A,C)}}^\mu =\left( 
\begin{array}{cc}
0 & \s^\mu\\ 
{\s^{\mu\; \#}} & 0
\end{array}
\right) ,\quad 
{{\overline \O}_{(A,C)}}^\mu =\left( 
\begin{array}{cc}
0 & {\tilde\s}^{\mu\; \#}\\ 
{\tilde \s^{\mu}} & 0
\end{array}
\right)  \
\end{eqnarray}
When the $B$ ($\equiv *$) structure is employed we have
\begin{eqnarray}
&& {\O_{(B)}}^\mu =\left( 
\begin{array}{cc}
0 & \s^\mu\\ 
{\tilde\s}^{\mu\; *} & 0
\end{array}
\right) ,\quad 
{{\overline \O}_{(B)}}^{\mu} =\left( 
\begin{array}{cc}
0 & {\s}^{\mu\; *}\\ 
{\tilde \s}^\mu & 0
\end{array}
\right)  
\end{eqnarray}
In this case
\begin{eqnarray}
{{\overline\Omega}_{(B)}}^{\mu} &=& {\Omega_{(B)}}^{\mu\; *}
\end{eqnarray} 
Due to the hermiticity property of the euclidean $\G$-matrices
in the euclidean space the $B$-structure and the 
$C$-structure coincide.\par
Let us furnish here for completeness the expression for $\O^{\mu\nu}$,
${\overline\O}^{\mu\nu}$ in the three $A$, $B$, $C$ cases. We get
\begin{equation}
{\O_{(A,C)}}^{\mu\nu} =\left( 
\begin{array}{cc}
\s^{\mu\nu}& 0\\ 
0 & -{{\tilde \s}}^{\mu\nu\; \#} 
\end{array}
\right) ,\quad 
{{\overline \O}_{(A,C)}}^{\mu\nu} =\left( 
\begin{array}{cc}
-{\s}^{\mu\nu \; \#} & 0\\ 
0 & {\tilde \s}^{\mu\nu } 
\end{array}
\right)  
\end{equation}
and respectively
\begin{equation}
{\O_{(B)}}^{\mu\nu} =\left( 
\begin{array}{cc}
\s^{\mu\nu}& 0\\ 
0 & {{\tilde \s}}^{\mu\nu\; *} 
\end{array}
\right) ,\quad 
{{\overline \O}_{(B)}}^{\mu\nu} =\left( 
\begin{array}{cc}
{\s}^{\mu\nu \; *} & 0\\ 
0 & {\tilde \s}^{\mu\nu} 
\end{array}
\right)  
\end{equation}
The solutionsof the free equations of motion in the $2D$- matrix formalism
(confront discussion at the end of section $2$) are expressed with the 
help of
$K = k_\mu {\O}^\mu$, ${\overline K} = k_\mu \O^\mu$ through
\begin{eqnarray}
k_\mu x^\mu \cdot \I &=& {1\over 2} ( K\cdot Z + {\overline Z} {\overline K})
\end{eqnarray}
so that 
\begin{eqnarray}
\partial_Z e^{i k_\mu x^{\mu} \cdot \I} &=& {i\over D} 
K e^{i k_\mu x^{\mu} \cdot \I}
\nonumber\\
\partial_{\overline Z} e^{i k_\mu x^{\mu} \cdot \I}
&=& {i\over D} {\overline K}e^{i k_\mu x^{\mu} \cdot \I}
\end{eqnarray}

\vspace{0.2cm}
\noindent{\section{A relativistic separation of variables.}}

Instead of using $Z$, ${\overline Z}$ we can make a change
of variables and introduce the $2D$-matrix coordinates 
$Z_\pm$ defined as follows:
\begin{eqnarray}
Z_\pm &=& Z\pm {\overline Z}
\label{zpm}
\end{eqnarray}
The commutativity property clearly still holds
\begin{eqnarray}
\relax [ Z_+, Z_-] &=& 0
\end{eqnarray}
while the $\partial_{\pm}$ matrix-derivatives can be introduced
\begin{eqnarray}
\partial_\pm &=& {1\over 2} (\partial_Z\pm \partial_{\overline Z} )
\end{eqnarray}
in order to satisfy, as a left action on $Z_\pm$,
\begin{eqnarray}
&& \partial_\pm Z_\pm = \I, \quad \partial_\pm Z_\mp = 0
\end{eqnarray}
The (pseudo)-euclidean quadratic form $ds^2\cdot\I$ can now be
read as follows
\begin{eqnarray}
d{\overline Z} \cdot dZ &=& {1\over 4} ( d{Z_+}^2 - d{Z_-}^2 )
\end{eqnarray}
Therefore $Z_+$ ($Z_-$) can be regarded as single-matrix coordinates, 
as those
introduced in section $2$, for the (pseudo)-euclidean spaces
associated (up to a global sign) to the quadratic forms ${dZ_+}^2$ 
and $d{Z_-}^2$ respectively.
In this context (\ref{zpm}) can be seen as a separation of 
variables which preserves the relativistic structure of the theory.\par
It is not difficult to prove that, while the $Z$, ${\overline Z}$ 
matrix-coordinates are
constructed with the full set of $x_\mu$ coordinates, no matter 
which structure
has been used to define them, $Z_+$ involves only half of the $x_\mu$
coordinates. The remaining ``half-sector" of the $x_\mu$'s appears 
in $Z_-$.
The even $D=(2n)$-dimensional spacetime is therefore splitted in two
$n$-dimensional relativistic spacetimes.\par
Such a result is a consequence of the following easy-to-prove 
equalities
\begin{eqnarray}
\O^{\mu} + {\overline \O}^{\mu} &=& \G^{\mu} +{\overline\G}^{\mu}
\nonumber\\
\O^{\mu} - {\overline \O}^{\mu} &=& \G^{D+1}\cdot 
(\G^{\mu} -{\overline\G}^{\mu})
\end{eqnarray}
where $\G^{D+1}$ has been introduced in (\ref{gammafivegen}). 
${\overline \G}^{\mu}$ denotes ${\G}^{\mu \; \dagger}$, ${\G}^{\mu\; T}$
or
${\G}^{\mu \; *}$ according to the context.\par
Let us analyze in detail the situation for each one of the three 
$A$, $B$, $C$ structures so far investigated.\par
The $A$ structure (the adjoint case) works only when the $D=2n$ 
spacetime
admits $n$ space directions
and $n$ time directions. We recall that the $\dagger$-conjugation 
property of 
$\G^{\mu}$ depends on its signature. It turns out as a consequence
that both $Z_+$, $Z_-$ describe an euclidean $n$-dimensional space 
(we apply on the space described by $Z_-$ an overall change of the 
signature).\par
For what concerns the $C$-structure we recall the results presented 
in the previous sections. 
We can denote as $t_+$ and $s_+$ the number of respectively timelike
and spacelike directions which are associated to symmetric $\G$-matrices.
Similarly $t_-$ and $s_-$ are the number of timelike and
spacelike directions whose $\G$-matrices are antisymmetric.
The relations ({\ref{ttss}) among $t_\pm$, $s_\pm$ hold.
As a consequence the $Z_+$ ($Z_-$) coordinate describes a relativistic 
spacetime
with signature $(t_+,s_+)$ (and respectively $(t_-,s_-)$).\par
The same result applies also when the $B$ (complex) structure is considered.
The vanishing of the index $I$ as introduced in (\ref{index}) now requires
$t_+=t_-$ and $s_+=s_-$. Let us $(t,s)\equiv (2k, 2n-2k)$ be the signature
of the original spacetime. The spacetime described by $Z_+$ results having
the same signature as the spacetime furnished by the $Z_-$ matrix coordinate
i.e.
\begin{eqnarray}
&&
(t_+,s_+) = (t_-,s_-) = (k,n-k).\nonumber
\end{eqnarray}
This is the last 
result which completes our analysis concerning the relativistic
separation of
variables.
 
\vspace{0.2cm}
\noindent{\section{The splitting case.}}

In this section we present a different way, alternative to the 
construction
so far employed, of solving the set of relations ({\ref{oobar}). It is
based on the $\G$-matrices realization expressed by the formula
(\ref{gammas}). Due to the presence in (\ref{gammas}) of tensor products 
of lower-dimensional 
spacetimes $\G$ matrices, the construction based on (\ref{gammas}) 
will be referred as the ``splitting case". It proceeds as follows.
At first we introduce two matrix-valued coordinates $X_+$ and $X_-$ 
through the positions
\begin{eqnarray}
X_+ &=& x_m \cdot \left( 
\begin{array}{cc}
0  & \I_q\otimes {\g_p}^{m} \\ 
\I_q\otimes {\g_p}^{m} 
& 0
\end{array}
\right) 
\label{xplus}
\end{eqnarray}
and 
\begin{eqnarray}
X_- &=& x_{\overline m} \cdot 
\left( 
\begin{array}{cc}
0  &  {\g_q}^{\overline m}\otimes \I_p \\ 
{\g_q}^{\overline m}\otimes \I_p  
& 0
\end{array}
\right) 
\label{xminus}
\end{eqnarray}
The conventions introduced in section $5$ are employed. In particular
$m$ takes value in a ($p+1$)-dimensional space and ${\overline m}$ in
a ($q+1$)-dimensional one. The total spacetime is $D=p+q+2$ 
(\ref{dimensionality}).\par
Clearly $X_\pm$ commute
\begin{eqnarray}
\relax [ X_+, X_-] &=& 0
\end{eqnarray}
The matrix coordinates $X_\pm$ realize a relativistic separation of
variables since the quadratic pseudoeuclidean form $ds^2$ can be written
as
\begin{eqnarray}
ds^2 \cdot \I &=& {dX_+}^2 + {dX_-}^2
\end{eqnarray}
The matrix derivatives $\partial_\pm$ can be introduced through
\begin{eqnarray}
\partial_+ &=& {1\over (p+1)} \cdot \partial_m
\left( 
\begin{array}{cc}
0  & \I_q\otimes {\g_p}^{m}\\ 
\I_q\otimes {\g_p}^{m}  
& 0
\end{array}
\right) 
\label{dplus}
\end{eqnarray}
and
\begin{eqnarray}
\partial_- &=&  {1\over (q+1)} \cdot\partial_{\overline m}
\left( 
\begin{array}{cc}
0  &  {\g_q}^{\overline m}\otimes \I_p \\ 
{\g_q}^{\overline m}\otimes \I_p  
& 0
\end{array}
\right) 
\label{dminus}
\end{eqnarray}  
$\partial_\pm$ are correctly normalized so that their left action 
on $X_\pm$ produce
\begin{eqnarray}
&& \partial_+ X_+ = \partial_- X_- = \I
\end{eqnarray}
Moreover the disentangling condition
\begin{eqnarray}
&& \partial_+ X_- =\partial_- X_+=0 
\end{eqnarray}
is verified.\par
With the help of $X_\pm$ we can construct the matrix-valued
$Z$, ${\overline Z}$ which allow to decompose the quadratic form
$ds^2$ through
\begin{eqnarray}
ds^2 \cdot \I &=& d{\overline Z}\cdot d Z
\nonumber
\end{eqnarray}
This can be done by setting
\begin{eqnarray}
Z &=& X_+ + i X_-\nonumber\\
{\overline Z} &=& X_+ - i X_-
\label{zzbarspl}
\end{eqnarray}
$Z$, ${\overline Z}$ commute. They can be regarded as $2D$ matrix-valued
coordinates as discussed in the introduction. In order to define a calculus
we need the introduction of the matrix derivatives $\partial_Z$, 
$\partial_{\overline Z}$. In the light of the ``splitting approach" here
discussed, this can be done unambiguously by setting
\begin{eqnarray}
\partial_Z &=& {1\over 2} (\partial_+ - i \partial_- )\nonumber\\
\partial_{\overline Z} &=& {1\over 2} ( \partial_+ + i \partial_- )
\end{eqnarray}
The above  $\partial_Z$, $\partial_{\overline Z}$ derivatives satisfy
all the required properties; they commute and moreover
\begin{eqnarray}
&&
\partial_Z Z =\partial_{\overline Z} {\overline Z} = \I\nonumber\\
&&
\partial_Z {\overline Z} = \partial_{\overline Z} Z = 0
\end{eqnarray}
as left action.\par
The only crucial point left is whether $\partial_Z$, 
$\partial_{\overline Z}$ realize a factorization of the d'Alembertian 
$\Box$ operator. It follows that
\begin{eqnarray}
\partial_{\overline Z} \partial_Z &=& {1\over 4} ({\partial_+}^2 +
{\partial_-}^2)
\end{eqnarray}
On the other hand ${\partial_\pm}^2$ satisfy
\begin{eqnarray}
{\partial_+}^2 &=& {1\over (p+1)^2} \I_q\otimes \I_p \Box_+\nonumber\\
{\partial_-}^2 &=& {1\over (q+1)^2} \I_q\otimes \I_p \Box_-
\end{eqnarray}
where $\Box_+$ and $\Box_-$ are the d'Alembertian for 
respectively the $(p+1)$ and the
$(q+1)$ dimensional subspaces. It turns
out that the property
\begin{eqnarray}
\partial_{\overline Z} \partial_Z &\propto  &\Box \nonumber
\end{eqnarray}
is verified only in the case 
\begin{eqnarray}
p &=& q
\end{eqnarray}
that is, the subspaces associated to $X_+$, $X_-$ have equal dimensions.\par
We recall that $p,q$ entering (\ref{gammas}) are even
dimensional, so that we can set $p=q=2n$. From the ({\ref{dimensionality})
condition $D=p+q+2 $, it follows that the $2D$-matrix coordinate calculus
can be introduced with the splitting condition only for spacetimes 
whose dimensionality $D$ is restricted to be an even integer of the kind
\begin{eqnarray}
D&=& 4 n +2
\end{eqnarray}
for some integral $n$.  
\par
This conclusion furnishes also the proof that the ``splitting case" here
considered is different from the previously analyzed $A$, $B$, $C$ cases.
In such cases only restrictions to the signature of the
spacetimes (for the $A$ and $B$ structures) were found, while
the dimensionality itself of the spacetimes is no further restricted
(besides the initial even-dimesionality requirement).\par
Let us conclude this section by pointing out that $Z$, ${\overline Z}$
can be represented as
\begin{eqnarray}
Z &=& x_\mu \O^\mu = 
x_m \O ^m + x_{\overline m} {\O}^{\overline m}\nonumber\\
{\overline Z} &=& 
x_\mu {\overline \O}^{\mu} 
= x_m {\O}^m + x_{\overline m} {\O}^{\overline m}  
\end{eqnarray}
($\mu$ is an index which spans the $D$ dimensional spacetime), where
$\O^{\mu}$, ${\overline \O}^{\mu}$
can be immediately read from (\ref{xplus},{\ref{xminus}) and 
(\ref{zzbarspl}).\par 
One can easily check that the derivatives $\partial_\pm$ of formulas
(\ref{dplus},\ref{dminus}) can be represented in the form
\begin{eqnarray}
\partial_Z &=& {1\over D} \partial_\mu {\overline \O}^{\mu}\nonumber\\
\partial_{\overline Z} &=& {1\over D} \partial_\mu {\O}^{\mu}
\end{eqnarray}
only when the equality $p=q$ is satisfied. The algebra which
has been analyzed in section $9$ can be formally recovered in the 
splitting ($S$) case. The ``splitting" is another construction which 
allows satisfying
the relations (\ref{oobar}).
 
\vspace{0.2cm}
\noindent{\section{An application to forms.}}

In the introduction we mentioned that one possible application for the
$2D$-matrix formalism consists in investigating abelian and Yang-Mills
gauge theories. In this respect it is convenient to outline how differential
forms can be introduced in the light of the $2D$ matrix formalism.
We sketch it here. Notations and conventions are those reported in section
$9$.\par
With the help of the differentials $d Z = dx_\mu \O^{\mu}$, 
$d{\overline Z} = dx_\mu {\overline \O}^{\mu}$ we can construct the 
wedge products
\begin{eqnarray}
dZ  \wedge dZ &=& {1\over 2} dx_\mu \wedge d x_\nu \cdot {\Xi_-}^{\mu\nu}
\nonumber
\\
d {\overline Z}  \wedge dZ &=& {1\over 2} dx_\mu \wedge d x_\nu \cdot
{{\overline\Omega}_-}^{\mu\nu}\nonumber
\\
dZ  \wedge d{\overline Z} &=& {1\over 2} dx_\mu \wedge d x_\nu\cdot 
{\Omega_-}^{\mu\nu}\nonumber
\\
d{\overline Z}  \wedge d
{\overline Z} &=& {1\over 2} dx_\mu \wedge d x_\nu\cdot 
{{\overline \Xi}_-}^{\mu\nu}
\end{eqnarray}
Notice that, due to the matrix character of the coordinates,
$dZ \wedge d Z \neq 0$ and similarly $d{\overline Z} \wedge d{\overline Z}
\neq 0$, while $dZ\wedge d{\overline Z} \neq - 
d{\overline Z} \wedge d Z$.\par
The differential operator $d$ which satisfies the nilpotency condition
\begin{eqnarray}
d^2 &=& 0
\end{eqnarray}
can be decomposed through 
\begin{eqnarray}
d\cdot \I &=& dx^\mu \partial_\mu \cdot \I = \partial + {\overline \partial}
\label{doperator}
\end{eqnarray}
where $\partial$, ${\overline \partial}$ are given by
\begin{eqnarray}
\partial &=& {D\over 2} \partial_Z \cdot d Z \nonumber \\
{\overline \partial} &=& {D\over 2} d {\overline Z} \cdot 
\partial_{\overline Z}
\end{eqnarray}
The equality (\ref{doperator}) is a consequence of the 
(\ref{oobar}) relation.\par
Please notice that the order in which derivatives and differentials
are taken is important because they are no longer commuting in the 
matrix case.\par
The wedge products between the differential operators $\partial$, 
${\overline \partial}$ is in general complicated. The simplest expression,
the only one which deserves being here reported, is for 
${\overline \partial}
\wedge \partial $:
\begin{eqnarray}
{\overline \partial} \wedge \partial &=& {1\over 32} \Box 
dx_\mu \wedge dx_\nu {\overline \O} ^{\mu\nu}
\end{eqnarray}
A one-form $A$ can be introduced with the positions
\begin{eqnarray}
A_Z &=& A_\mu {\overline \O}^\mu\nonumber\\
A_{\overline Z} &=& A_{\mu} \O^\mu
\end{eqnarray}
Indeed we have for $A$
\begin{eqnarray}
A &=& A_\mu dx^\mu \cdot \I = {1\over 2} 
( A_Z dZ + d{\overline Z} A_{\overline Z} ) 
\end{eqnarray}
In the abelian case a gauge transformation is simply
realized by the mapping 
\begin{eqnarray}
&&
A \mapsto A + d \Lambda = A
+ (\partial +{\overline\partial}) \Lambda \nonumber
\end{eqnarray}
where $\Lambda$
is a matrix-valued $0$-form.\par
The stress-energy tensors $F_{\mu\nu}$ are introduced as 
two-forms with the standard procedure
\begin{eqnarray}
F &=& dx_{\mu}\wedge dx_\nu F^{\mu\nu}\cdot \I =
{1\over 2} dx_\mu \wedge dx_\nu (\partial^\mu A^\nu -
\partial^\nu A^\mu ) \cdot \I =
\nonumber\\
&=& d A = 
 (\partial + {\overline \partial} ) A
\end{eqnarray}
 
\vspace{0.2cm}
\noindent{\section{Conclusions.}}

In this paper we have introduced a matrix-calculus to describe relativistic
field theories in higher-dimensional spacetimes. We discussed the
single-matrix approach, which can be applied for instance to scalar
bosonic theories, and the $2D$ matrix calculus, by far more
general, which employes matrix-valued $Z$, ${\overline {Z}}$ coordinates. We
pointed out the manifest Lorentz-covariance of our approach; furthermore
we investigated the consistency conditions which made it possible.\par 
In order to solve this problem we produced some other results as byproducts.
The recursive formula (\ref{gammas}) to construct 
$\G$-matrices is an example.
The computation of the coefficients in 
the ``vector-trace" formulas 
(\ref{vectortraces}) is another one. 
This computation has also lead us to introduce 
an index labelling 
inequivalent $\G$-structures under conjugation realized by
real orthogonal matrices. Such an index describes as well the equivalence 
classes of Wick rotations from the euclidean into the pseudoeuclidean 
spacetimes.\par
Since a short summary of the main results here presented has already been 
furnished in the introduction, we do not repeat it now. Rather, we prefer
to give some commentaries concerning the potentialities of the formalism
we have constructed. \par
It surely deserves being stressed the fact that the existence of a $2D$ 
matrix-calculus relies on non-trivial properties concerning dimension and
signature of spacetimes. These properties are described by nice mathematical 
formulas. At a purely formal level we dispose of a very attractive 
mathematical construction. Spacetimes of different dimensions and signature
can be formally treated on equal footing. The different properties they
share are automatically encoded in the calculus. This feature could be even
more relevant for its supersymmetric extension (presently under 
construction). It is expected to put even more restrictions on the allowed
spacetimes. It seems more than a mere possibility that the spacetimes 
which can be consistently defined would be those obtained from superstring.
The question concerning the nature of the spacetime and its signature
\cite{hull} can in principle be raised for the $2D$ matrix calculus.\par     
At a less formal level and more down-to-earth point of view, we have of
course to ask
ourselves the question about the usefulness and applicability of the whole
construction. So, let us state it clearly. We dispose of a formalism which
can be jokingly named as ``fat flat space" (where ``fat" stands for matrix).
In the present paper we have just unveiled the basic roots of such a
formalism. Of course more work is required to introduce e.g. lagrangians, 
Poisson brackets, hamiltonians and so on, or to deal with curved
spacetimes, but in fact there is no 
obstacle in performing such extensions. Indeed they can be carried out
quite straightforwardly. The main point here is that our construction
can in principle lead to investigate higher-dimensional relativistic field
theories by borrowing the techniques employed for standard $2D$ physics.\par
In the introduction we already mentioned the issue of integrability. In fact
we have a lot more. In standard $2D$ physics hamiltonian methods are widely
used. They are more powerful than lagrangian methods and, due to the fact
that the $2D$ Poincar\'e invariance admits only three generators, in just 
$2D$ the loss of the manifest Lorentz-covariance implicit in the hamiltonian 
approach is not a such a big loss. Our $Z$, ${\overline {Z}}$ coordinates can in principle
be used for such a hamiltonian description. Moreover, issues like current
algebras can be investigated in the light of the $2D$ matrix approach.
This could mean the extension of WZNW theories to higher-dimensional
spacetimes (see \cite{genwznw}), as well as their possible hamiltonian
reductions (\cite{toda}) to higher-dimensional relativistic Toda field 
theories (\cite{gentoda}).\par
Another topic in mathematical physics which can profit of the present
formalism concerns issues of index theorem and computation of the index 
for elliptic operators in higher dimension. The recursive formula
(\ref{gammas}) provides the basis for factorizing elliptic operators 
(just repeating the steps done for the standard d'Alembertian). Heat-kernel
computations can be made in terms of the $2D$-matrix calculus.\par
Let us finally mention that the ``splitting of variables" described
in section $11$ admits an useful application in analyzing 
reductions from higher-dimensional spacetime to lower-dimensional ones,   
with a procedure which can be regarded as a ``folding" of spacetimes
(allowing to express e.g. de Sitter or anti-de Sitter spacetimes from
an underlying $10$-dimensional theory). This is the content of a work 
currently at an advanced stage of preparation.

\vspace{0.2cm}
\par
\noindent{\Large{\bf{Acknowledgments}}}
\quad\par
{\quad}
\par
We are grateful to CBPF, where this work has been elaborated,
for the kind hospitality. It is a real pleasure to thank J. A. Helayel 
Neto for both stimulating discussions and the warm and friendly 
atmosphere we enjoyed at DCP.
\par

\end{document}